\documentclass{article}
\usepackage[T1]{fontenc}
\usepackage[latin1]{inputenc}
\usepackage{array}
\usepackage{float}
\usepackage{graphicx}
\IfFileExists{url.sty}{\usepackage{url}}
                      {\newcommand{\url}{\texttt}}

\makeatletter

\providecommand{\tabularnewline}{\\}

\makeatother
\begin{document}

\title{FDCSUSYDecay:
A MSSM Decay Package}

\author{Wei Qi, J.X. Wang\\
\\
\textit{Institute of High Energy Physics, the Chinese Academy of Sciences},\\
\textit{P.O.Box 918(4), Beijing 100049, China}
}

\maketitle
\begin{abstract}
FDCSUSYDecay is a FORTRAN program package generated by FDC (Feynman
Diagram Calculation) system fully automatically. It is dedicated to
calculate at tree-level all the possible 2-body decays of
SUSY and Higgs particles in the Minimal Supersymmetric extension of the 
Standard Model (MSSM). The format of its output files complies
with SUSY Les Houches Accord and can
be easily imported by other packages.

\begin{flushleft}


\end{flushleft}
\end{abstract}

\textbf{\Large Program Summary}{\Large \par}

\emph{Program Title:} FDCSUSYDecay (Version 1.00)

\emph{Journal Reference:}

\emph{Catalogue identifier:}



\emph{Operating system:} Linux

\emph{Programming language:} FORTRAN 77

\emph{External libraries:} CERNLIB 2003 (or up)

\emph{Distribution format:} tar gzip file

\emph{Size of the compressed distribution file:} 622,560 Bytes.

\emph{Classification:} 11.1

\emph{Nature of physical problem:} This package can calculate all
the possible SUSY particle and Higgs 2-body decay width and branch
ratio at tree-level in the MSSM model.

\emph{Method of solution:} 
By running FDC, the Feynman rules for the MSSM model are generated,
all the decay widths are calculated analytically and corresponding
FORTRAN codes are generated for this package.

\section{Generality}
With the progress in high energy physics, it is obvious that an
automatic calculation of the physical processes by the perturbative
quantum field theory becomes very useful and important.  There
are many projects for this purpose such as \texttt{CompHEP}\cite{comphep1995},
\texttt{Grace}\cite{grace1993} and so on. FDC (Feynman Diagram Calculation)
\cite{Wang:2004du} is one of such kind projects. It is mainly
written in REDUCE and RLisp. As designed to be, it can conveniently
implement physical models from the first principle, then construct their 
\texttt{\emph{Lagrangian}}s, deduce the Feynman rules and generate
FORTRAN codes for calculations of physical quantities such as decay
width, cross-section and matrix-element event generator. 
The MSSM model is implemented in FDC\cite{jp2000} and the generated 
results could be found on FDC homepage\footnote{\url{http://www.ihep.ac.cn/lunwen/wjx/public_html}}. To check the results such as the 
mixing matrices, Feynman rules, as well as to provide an application of FDC,
we developed a program to automatically generate the FORTRAN packages 
\texttt{FDCSUSYDecay} calculating SUSY particle decays and 
\texttt{FDCEvent@LHC} \cite{fdcevent_lhc} which is a  
matrix-element event generator of the MSSM related processes at LHC. 

There are many program packages which can calculate the SUSY particle
decays, such as multi-purpose packages PYTHIA, ISAJET and SUSYGEN or
dedicated packages HDecay (which is special for Higgs decay in the SM and
MSSM) \cite{hdecay} and SDecay \cite{sdecay}, but the codes of these
packages are manually typed. However, FDCSUSYDecay is a FORTRAN
program package which is fully automatically generated by FDC. 
In this way many uncareful errors can be systematically debugged because
these artificial errors in the model would affect a series of the resulting Feynman
vertices and thus the corresponding decay equations which could be found easily.
All the possible 2-body decay processes (in total 543 processes) for SUSY 
and Higgs particles in the MSSM model have been included. 
This package can calculate the decay width and branch ratio at 
Snowmass benchmark points 1-9 \cite{sps} (Tab. \ref{hepar}) or
any user-specified parameter configration.
The format of its output files complies with the SUSY Les
Houches Accord (SLHA)\cite{slha} and thus can be imported by other
programs conveniently. 

Although presently only 2-body SUSY decays are implemented in this
package, the FORTRAN codes for any 3-body decay can 
be generated by FDC. Since 
the important decay channels can not be distinguished   
from neglectable ones before the MSSM parameters are chosen and to include
all 3-body decays will make the length of the generated FORTRAN
codes out of control, the 3-body decays are not included in the package. 
Alternatively, The package with needed 3-body channels can be generated by 
FDC when the MSSM parameters fixed.  

The package can be used in two modes. The first mode is called the high-energy scale mode. In this mode, users need to input the parameters in high-energy scale 
scenarios (mSUGRA, GMSB and AMSB) or choose from nine Snowmass benchmark 
points (which have been fed into this package hard-wiredly). Then the program 
obtains all the parameters at SUSY-breaking scale through interface 
with ISAJET \cite{isajet} and then calculates everything such as masses, 
mixing matrices, coupling constants and decay width. The second mode is called 
low-energy scale mode, in which users need to input manually the MSSM parameters:
\begin{itemize}
\item {\footnotesize $M_{1},M_{2},M_{3}$ : U(1), U(2), U(3) gaugino mass
parameters}{\footnotesize \par}
\item {\footnotesize $\mu$ : Higgsino mass parameter}{\footnotesize \par}
\item {\footnotesize $\tan\beta$ : the ratio of the} \emph{\footnotesize vev's}
{\footnotesize of the two-Higgs doublet fields}{\footnotesize \par}
\item {\footnotesize $A_{\tau},A_{b},A_{t}$ : the 3rd generation trilinear
couplings}{\footnotesize \par}
\item {\footnotesize $m_{A^{0}}$ : the neutral pseudoscalar Higgs ($A^{0}$)
mass}{\footnotesize \par}
\item {\footnotesize $\tilde{m}_{\tilde{e}_{L}}^{2}(=\tilde{m}_{\tilde{\nu}_{eL}}^{2}),\tilde{m}_{\tilde{e}_{R}}^{2},\tilde{m}_{\tilde{\mu}_{L}}^{2}(=\tilde{m}_{\tilde{\nu}_{\mu L}}^{2}),\tilde{m}_{\tilde{\mu}_{R}}^{2},$
}\\
{\footnotesize $\tilde{m}_{\tilde{u}_{L}}^{2}(=\tilde{m}_{\tilde{d}_{L}}^{2}),\tilde{m}_{\tilde{u}_{R}}^{2},\tilde{m}_{\tilde{d}_{R}}^{2},\tilde{m}_{\tilde{c_{L}}}^{2}(=\tilde{m}_{\tilde{s}_{L}}^{2}),\tilde{m}_{\tilde{c}_{R}}^{2},\tilde{m}_{\tilde{s}_{R}}^{2}$
: the SUSY-breaking mass parameters for the 1st/2nd generation sfermions}{\footnotesize \par}
\item {\footnotesize $\tilde{m}_{\tilde{\tau}_{1}}^{2},\tilde{m}_{\tilde{\tau}_{2}}^{2},\tilde{m}_{\tilde{b}_{2}}^{2},\tilde{m}_{\tilde{t}_{1}}^{2}(=\tilde{m}_{\tilde{b}_{1}}^{2}),\tilde{m}_{\tilde{t}_{2}}^{2}$
: the SUSY-breaking mass parameters for the 3rd generation sfermions}{\footnotesize \par}
\end{itemize}
The Yukawa couplings for the 1st and 2nd generation fermions are fixed to zero as an approximation and the corresponding soft SUSY breaking terms are also fixed to zero in the MSSM model used here. If needed, they can be easily included in our package by FDC.
\begin{center}
\begin{table}
\begin{tabular}{cm{12em}ccccc}
\hline 
{\footnotesize SPS}&
{\footnotesize mSUGRA scenario}&
{\footnotesize $m_{0}$}&
{\footnotesize $m_{1/2}$}&
{\footnotesize $A_{0}$}&
{\footnotesize $\tan\beta$}&
{\footnotesize sign $\mu$}\tabularnewline
\hline
{\footnotesize 1}&
{\footnotesize typical point}&
{\footnotesize 100}&
{\footnotesize 250}&
{\footnotesize -100}&
{\footnotesize 10}&
{\footnotesize +}\tabularnewline
{\footnotesize 2}&
{\footnotesize focus point region}&
{\footnotesize 1450}&
{\footnotesize 300}&
{\footnotesize 0}&
{\footnotesize 10}&
{\footnotesize +}\tabularnewline
{\footnotesize 3}&
{\footnotesize model line into coannihilation region}&
{\footnotesize 90}&
{\footnotesize 400}&
{\footnotesize 0}&
{\footnotesize 10}&
{\footnotesize +}\tabularnewline
{\footnotesize 4}&
{\footnotesize large $\tan\beta$}&
{\footnotesize 400}&
{\footnotesize 300}&
{\footnotesize 0}&
{\footnotesize 50}&
{\footnotesize +}\tabularnewline
{\footnotesize 5}&
{\footnotesize light stop}&
{\footnotesize 150}&
{\footnotesize 300}&
{\footnotesize -1000}&
{\footnotesize 5}&
{\footnotesize +}\tabularnewline
{\footnotesize 6}&
{\footnotesize non-unified gaugino masses $M_{1}=480,M_{2}=M_{3}=300$}&
{\footnotesize 150}&
{\footnotesize 300}&
{\footnotesize 0}&
{\footnotesize 10}&
{\footnotesize +}\tabularnewline
\hline 
&
{\footnotesize GMSB scenario}&
{\footnotesize $\Lambda$}&
{\footnotesize $M_{mess}$}&
{\footnotesize $N_{mess}$}&
{\footnotesize $\tan\beta$}&
{\footnotesize sign $\mu$}\tabularnewline
\hline
{\footnotesize 7}&
{\footnotesize NLSP = $\tilde{\tau_{1}}$}&
{\footnotesize 40,000}&
{\footnotesize 80,000}&
{\footnotesize 3}&
{\footnotesize 15}&
{\footnotesize +}\tabularnewline
{\footnotesize 8}&
{\footnotesize NLSP = $\tilde{\chi_{1}^{0}}$}&
{\footnotesize 100,000}&
{\footnotesize 200,000}&
{\footnotesize 1}&
{\footnotesize 15}&
{\footnotesize +}\tabularnewline
\hline 
&
{\footnotesize AMSB scenario}&
{\footnotesize $m_{0}$}&
{\footnotesize $m_{3/2}$}&
&
{\footnotesize $\tan\beta$}&
{\footnotesize sign $\mu$}\tabularnewline
\hline
{\footnotesize 9}&
{\footnotesize small $\Delta m(\tilde{\chi_{1}^{+}-\tilde{\chi_{1}^{0})}}$}&
{\footnotesize 400}&
{\footnotesize 60,000}&
&
{\footnotesize 10}&
{\footnotesize +}\tabularnewline
\hline
\end{tabular}
\caption{High-energy scale scenario points corresponding to
Snowmass points 1-9\cite{compilation}
\label{hepar}}
\end{table}
\end{center}

\section{Installation and Usage}
\subsection{Installation}
The package is distributed in a gzip compressed file named \emph{fsdecay.tar.gz} and can be downloaded from FDC homepage:
\begin{quote}
\url{http://www.ihep.ac.cn/lunwen/wjx/public_html}
\end{quote}
The CERN program library (CERNLIB) 2003 \cite{cernlib} (or up) and ISAJET are required only for high-energy scale mode.
The installation steps are given below and for convenience it is assumed 
that\footnote{Users can substitute the user account and paths for their owns, when installing on their own computers.}:

\begin{enumerate}
\item CERNLIB is located at directory \emph{/cern/2003},
\item the current user account is \emph{linuxuser;}
\item the downloaded file \emph{fsdecay.tar.gz} are stored at directory \emph{\textasciitilde{}linuxuser;}
\item the package will be installed to
\emph{\textasciitilde{}linuxuser/FDCDecay}.
\end{enumerate}
The installation steps are 
\begin{enumerate}
\item Create the target directory and go into it by
\begin{quote}
\texttt{mkdir \textasciitilde{}linuxuser/FDCDecay}\\
\texttt{cd \textasciitilde{}linuxuser/FDCDecay}
\end{quote}
\item Decompress the downloaded file by
\begin{quote}
\texttt{tar -xzvf \textasciitilde{}linuxuser/fsdecay.tar.gz}
\end{quote}
\item Specify the CERNLIB path by adding following line to the \textbf{very beginning} of the file \emph{makefile}\footnote{This step is necessary only for high-energy scale mode.}
\begin{quote}
\texttt{cern = /cern/2003}
\end{quote}
\item Compile the low-energy scale mode by
\begin{quote}
\texttt{make lowscale}
\end{quote}
\item Compile the high-energy scale mode by
\begin{quote}
\texttt{make highscale}
\end{quote}
\item Delete the temporary files by
\begin{quote}
\texttt{make clean}
\end{quote}
\end{enumerate}
After steps above without encountering any errors, one will obtain two
executable files named \emph{fsdecayhigh} and \emph{fsdecaylow} in
directory \emph{\textasciitilde{}linuxuser/FDCDecay} which are 
stand-alone programs used to calculate decay width and branch ratio
in high-energy scale and low-energy scale modes respectively.
\subsection{Usage}
As mentioned above, with this package, users can calculate decays
in high-energy scale mode or low-energy scale mode. While in 
high-energy scale mode, users can either specify the scenarios
(including mSUGRA, GMSB, AMSB) and their parameters manually or just use
the Snowmass benchmark points. While in low-energy scale
mode, users must input the MSSM (and the Standard Model) parameters
manually. 

For this purpose, we prepared two executable programs \emph{fsdecayhigh}
and \emph{fsdecaylow} corresponding to these two modes respectively. 

Next we will describe the usage of these two programs. Test-run examples
are given in Sec. \ref{sec:Test-Run}.
\subsubsection{Using high-energy scale mode}
It is very simple to calculate decay width in high-energy scale
mode. What users need to do is just run program \emph{fsdecayhigh}
and then the program will give prompts on screen and require corresponding
inputs.
\subsubsection{Using low-energy scale mode}

To calculate decay width in this mode is somewhat more complicated.
Users should follow steps below:

\begin{enumerate}
\item Modify the FORTRAN file \emph{usrpara.f} in the installation directory
to assign value to parameters.
There is a subroutine \emph{FUPAR} in which users can specify the MSSM and 
SM parameters respectively
by assigning values to corresponding variables. All these variables
are listed in Tab.\ref{tab:mssmandsmparvar}.
\item Re-build the executable program \emph{fsdecaylow} by command:\emph{}\\
\emph{make lowscale clean}
\item Run program \emph{fsdecaylow} and it will give prompt for users
to specify the output file name
\end{enumerate}
\textbf{\Large }%
\begin{table}
\centering
\begin{tabular}{cc|cc||cc}
\hline 
\multicolumn{4}{c||}{MSSM parameters}&
\multicolumn{2}{c}{SM parameters}\tabularnewline
\hline 
\texttt{\footnotesize FSSGM1}&
{\footnotesize $M_{1}$}&
\texttt{\footnotesize FSSSBMUR}&
{\footnotesize $\tilde{m}_{\tilde{\mu}_{R}}^{2}$}&
\texttt{\footnotesize FSMME}&
{\footnotesize $m_{e}$}\tabularnewline
\texttt{\footnotesize FSSGM2}&
{\footnotesize $M_{2}$}&
\texttt{\footnotesize FSSSBTAU1}&
{\footnotesize $\tilde{m}_{\tilde{\tau}_{1}}^{2}$}&
\texttt{\footnotesize FSMMU}&
{\footnotesize $m_{\mu}$}\tabularnewline
\texttt{\footnotesize FSSGM3}&
{\footnotesize $M_{3}$}&
\texttt{\footnotesize FSSSBTAU2}&
{\footnotesize $\tilde{m}_{\tilde{\tau}_{2}}^{2}$}&
\texttt{\footnotesize FSMMTAU}&
{\footnotesize $m_{\tau}$}\tabularnewline
\texttt{\footnotesize FSSHMU}&
{\footnotesize $\mu$}&
\texttt{\footnotesize FSSSBUPL}&
{\footnotesize $\tilde{m}_{\tilde{u}_{L}}^{2}$}&
\texttt{\footnotesize FSMMUP}&
{\footnotesize $m_{u}$}\tabularnewline
\texttt{\footnotesize FSSTANB}&
{\footnotesize $\tan\beta$}&
\texttt{\footnotesize FSSSBUPR}&
{\footnotesize $\tilde{m}_{\tilde{u}_{R}}^{2}$}&
\texttt{\footnotesize FSMMDN}&
{\footnotesize $m_{d}$}\tabularnewline
\texttt{\footnotesize FSSMHA}&
{\footnotesize $m_{A^{0}}$}&
\texttt{\footnotesize FSSSBCHL}&
{\footnotesize $\tilde{m}_{\tilde{c}_{L}}^{2}$}&
\texttt{\footnotesize FSMMCH}&
{\footnotesize $m_{c}$}\tabularnewline
\texttt{\footnotesize FSSAAL}&
{\footnotesize $A_{\tau}$}&
\texttt{\footnotesize FSSSBCHR}&
{\footnotesize $\tilde{m}_{\tilde{c}_{R}}^{2}$}&
\texttt{\footnotesize FSMMST}&
{\footnotesize $m_{s}$}\tabularnewline
\texttt{\footnotesize FSSAAT}&
{\footnotesize $A_{t}$}&
\texttt{\footnotesize FSSSBDNR}&
{\footnotesize $\tilde{m}_{\tilde{d}_{R}}^{2}$}&
\texttt{\footnotesize FSMMTP}&
{\footnotesize $m_{t}$}\tabularnewline
\texttt{\footnotesize FSSAAB}&
{\footnotesize $A_{b}$}&
\texttt{\footnotesize FSSSBSTR}&
{\footnotesize $\tilde{m}_{\tilde{s}_{R}}^{2}$}&
\texttt{\footnotesize FSMMBT}&
{\footnotesize $m_{b}$}\tabularnewline
\texttt{\footnotesize FSSSBEL}&
{\footnotesize $\tilde{m}_{\tilde{e}_{L}}^{2}$}&
\texttt{\footnotesize FSSSBBT2}&
{\footnotesize $\tilde{m}_{\tilde{b}_{2}}^{2}$}&
\texttt{\footnotesize FSMMZ}&
{\footnotesize $m_{z}$}\tabularnewline
\texttt{\footnotesize FSSSBER}&
{\footnotesize $\tilde{m}_{\tilde{e}_{R}}^{2}$}&
\texttt{\footnotesize FSSSBTP1}&
{\footnotesize $\tilde{m}_{\tilde{t}_{1}}^{2}$}&
\texttt{\footnotesize FSMSN2THW}&
{\footnotesize $\sin^{2}(\theta_{W})$}\tabularnewline
\texttt{\footnotesize FSSSBMURL}&
{\footnotesize $\tilde{m}_{\tilde{\mu}_{L}}^{2}$}&
\texttt{\footnotesize FSSSBTP2}&
{\footnotesize $\tilde{m}_{\tilde{t}_{2}}^{2}$}&
\texttt{\footnotesize FSMFNST}&
{\footnotesize $\alpha_{EM}$}\tabularnewline
&
&
&
&
\texttt{\footnotesize FSMALPS}&
{\footnotesize $\alpha_{s}$}\tabularnewline
\hline
\end{tabular}{\scriptsize \par}
\caption{Variables corresponding to the MSSM and SM parameters in subroutine \emph{FUPAR}\label{tab:mssmandsmparvar}}
\end{table}

\section{Output files}
The format of the generated output file for decays complies with the SUSY Les Houches
Accord\cite{slha}. It is composed of several blocks
which are separated by empty lines. Each of these blocks describes
a SUSY particle decay. Every block begins with a \texttt{DECAY} statement
specifying the decaying mother particle and its total width, in the
format as follows
{\footnotesize
\begin{verbatim}
#       PDG         WIDTH(GeV)
DECAY   2000006     9.00247665E+00   # ~t_2 decays              
\end{verbatim}
}
The first integer is the PDG particle code\cite{PDG2004} of the decaying
particle. The subsequent real number is its total width.
The end comment gives a human readable translation of the PDG code.
All the decay modes for this mother particle are listed in the subsequent lines in the format
{\footnotesize
\begin{verbatim}
#    BR                NDA   ID1       ID2
     2.25878090E-01    2          23   1000006   # ~t_2 -> Z0  ~t_1         
     2.21288383E-01    2           5   1000037   # ~t_2 -> b  ~chi+_2       
     1.90569923E-01    2           6   1000035   # ~t_2 -> t  ~chi0_4       
     1.61424758E-01    2           5   1000024   # ~t_2 -> b  ~chi+_1       
  ...     
\end{verbatim}
}
In each line above, the first real number is the branch ratio of
the decay mode, the first integer is the number of daughters (which
is always 2 in this package) and the following integers are the PDG
codes of these daughters. The specific FORTRAN format for the \texttt{DECAY}
statement and the entries in a block are respectively
{\footnotesize
\begin{verbatim}
      FORMAT('#',7X,'PDG',9X,'WIDTH(GeV)')
      FORMAT('DECAY',1X,I9,3X,1P,E16.8,0P,3X,'#',1X,A25)
      FORMAT('#',4X,'BR',16X,'NDA',3X,'ID1',7X,'ID2')
      FORMAT(3X,1P,E16.8,0P,3X,I2,3X,2(I9,1X),2X,'#',1X,A25)
\end{verbatim}
}

\section{Cross-Check with ISAJET}
\subsection{Running mass parameters and higher-order mass corrections}
In ISAJET (Version 7.64), running mass for $\tau$, $b$ and $t$ are used in the
calculations of stau($\tilde{\tau}_{i}$), sbottom ($\tilde{b}_{i}$)
and stop ($\tilde{t}_{i}$) mass eigenvalues, mixing angles and in
the vertices of $fermion(\tau/b/t)\sim sfermion(\tilde{\tau}_{i}/\tilde{b}_{i}/\tilde{t}_{i})\sim gaugino(\tilde{\chi}_{i}^{\pm}/\tilde{\chi}_{i}^{0})$. 

Higher order corrections are added in ISAJET to the mass of Higgs
$H^{0}$, $h^{0}$, $H^{+}$ and gluino $\tilde{g}$. The corrections
at SPS 1-9 bring less than $1\%$ modification to the $H^{0}$ and $H^{+}$
mass, $27\%\sim43\%$ to $h^{0}$ mass and $5\%\sim15\%$ for $\tilde{g}$
mass. In Tab.\ref{masscorr} we listed the Higgs and gluino mass at
SPS 1-9 from ISAJET and our decay package. However these corrections
have not been implemented in FDC yet (thus in our package). 

In order to compare the decay widths, we take, \textbf{temporarily},
the running mass and the corrected Higgs and gluino mass as inputs
from ISAJET.

\begin{table}
 \centering
{\scriptsize }\begin{tabular}{|c|cc|cc|cc|cc|}
\hline 
{\scriptsize SPS }&
 {\scriptsize $M_{H^{0}}^{I}$}&
 {\scriptsize $M_{H_{0}}^{F}$}&
 {\scriptsize $M_{h^{0}}^{I}$}&
 {\scriptsize $M_{h^{0}}^{F}$}&
 {\scriptsize $M_{H^{+}}^{I}$}&
 {\scriptsize $M_{H^{+}}^{F}$}&
 {\scriptsize $M_{\tilde{g}}^{I}$}&
 {\scriptsize $M_{\tilde{g}}^{F}$}\tabularnewline
\hline
{\scriptsize 1 }&
 {\scriptsize 395.629 }&
 {\scriptsize 395.310 }&
 {\scriptsize 113.657 }&
 {\scriptsize 89.266 }&
 {\scriptsize 403.229 }&
 {\scriptsize 402.885 }&
 {\scriptsize 606.633 }&
 {\scriptsize 579.093 }\tabularnewline
 {\scriptsize 2 }&
 {\scriptsize 1501.688 }&
 {\scriptsize 1499.753 }&
 {\scriptsize 115.551 }&
 {\scriptsize 89.358 }&
 {\scriptsize 1503.698 }&
 {\scriptsize 1501.773 }&
 {\scriptsize 794.334 }&
 {\scriptsize 691.835 }\tabularnewline
 {\scriptsize 3 }&
 {\scriptsize 581.559 }&
 {\scriptsize 580.898 }&
 {\scriptsize 116.663 }&
 {\scriptsize 89.320 }&
 {\scriptsize 586.758 }&
 {\scriptsize 586.088 }&
 {\scriptsize 932.097 }&
 {\scriptsize 893.408 }\tabularnewline
 {\scriptsize 4 }&
 {\scriptsize 356.491 }&
 {\scriptsize 356.153 }&
 {\scriptsize 114.965 }&
 {\scriptsize 91.092 }&
 {\scriptsize 369.506 }&
 {\scriptsize 364.995 }&
 {\scriptsize 732.394 }&
 {\scriptsize 686.647 }\tabularnewline
 {\scriptsize 5 }&
 {\scriptsize 690.826 }&
 {\scriptsize 691.094 }&
 {\scriptsize 119.901 }&
 {\scriptsize 84.047 }&
 {\scriptsize 695.257 }&
 {\scriptsize 694.805 }&
 {\scriptsize 719.234 }&
 {\scriptsize 687.721 }\tabularnewline
 {\scriptsize 6 }&
 {\scriptsize 464.004 }&
 {\scriptsize 463.524 }&
 {\scriptsize 114.433 }&
 {\scriptsize 89.294 }&
 {\scriptsize 470.489 }&
 {\scriptsize 470.008 }&
 {\scriptsize 717.950 }&
 {\scriptsize 684.761 }\tabularnewline
 {\scriptsize 7 }&
 {\scriptsize 390.072 }&
 {\scriptsize 389.560 }&
 {\scriptsize 113.537 }&
 {\scriptsize 90.317 }&
 {\scriptsize 398.160 }&
 {\scriptsize 397.483 }&
 {\scriptsize 943.832 }&
 {\scriptsize 894.279 }\tabularnewline
 {\scriptsize 8 }&
 {\scriptsize 532.413 }&
 {\scriptsize 531.719 }&
 {\scriptsize 114.802 }&
 {\scriptsize 90.339 }&
 {\scriptsize 538.381 }&
 {\scriptsize 537.554 }&
 {\scriptsize 835.300 }&
 {\scriptsize 753.465 }\tabularnewline
 {\scriptsize 9 }&
 {\scriptsize 926.592 }&
 {\scriptsize 925.408 }&
 {\scriptsize 115.089 }&
 {\scriptsize 89.347 }&
 {\scriptsize 929.825 }&
 {\scriptsize 928.678 }&
 {\scriptsize 1294.129 }&
 {\scriptsize 1225.333  }\tabularnewline
\hline
\end{tabular}{\scriptsize \par}
\caption{Higgs and gluino mass in ISAJET (I) and FDCSUSYDecay (F)\label{masscorr}}
\end{table}

\subsection{Cross-check and results}
Taking higher order corrections to Higgs and gluino mass as inputs,
we calculated the decay widths at SPS 1-9 and compared them with the
data from ISAJET.

For the decay modes with $10\%$ discrepancies or larger between ISAJET
and our package, we analyzed the decay equations in the source codes
of ISAJET (Version 7.64). Two reasons for the discrepancies were found: (1) correction
to triple Higgs vertices and QCD radiative correction to $Higgs\rightarrow q(t/b)+\overline{q}(\overline{t}/\overline{b})$
included in ISAJET; (2) different decay equations used by ISAJET and
our package which imply something wrong in the codes of ISAJET\footnote{We carefully checked the MSSM model used in FDC with that in Ref. \cite{Kuroda:1999ks}}. 

For the first kind of discrepancy, we list the widths of Higgs decays 
which are influenced by the higher order
corrections in Tab.\ref{tblhiggsdecay}. From Tab.\ref{tblhiggsdecay} we can 
see that the corrections bring large modifications sometimes.

\begin{table}
{\scriptsize }\begin{tabular}{|c|c|c||c|c|c||ccc|}
\hline 
\multicolumn{3}{|c||}{{\scriptsize $H^{0}\rightarrow b+\overline{b}$}}&
\multicolumn{3}{c||}{{\scriptsize $H^{0}\rightarrow t+\overline{t}$}}&
\multicolumn{3}{c|}{{\scriptsize $H^{0}\rightarrow h^{0}+h^{0}$}}\tabularnewline
\hline
{\scriptsize SPS }&
 {\scriptsize $\Gamma_{I}$}&
 {\scriptsize $\Gamma_{F}$}&
{\scriptsize SPS }&
 {\scriptsize $\Gamma_{I}$}&
 {\scriptsize $\Gamma_{F}$}&
{\scriptsize SPS }&
 {\scriptsize $\Gamma_{I}$}&
 {\scriptsize $\Gamma_{F}$}\tabularnewline
\hline
{\scriptsize 1 }&
 {\scriptsize 1.131 }&
 {\scriptsize 1.956 }&
{\scriptsize 1 }&
 {\scriptsize 0.0435 }&
 {\scriptsize 0.0300 }&
{\scriptsize 1 }&
 {\scriptsize 0.00120 }&
 {\scriptsize 0.00945 }\tabularnewline
 {\scriptsize 2 }&
 {\scriptsize 3.621 }&
 {\scriptsize 7.446 }&
 {\scriptsize 2 }&
 {\scriptsize 0.779 }&
 {\scriptsize 0.851 }&
 {\scriptsize 2 }&
 {\scriptsize 0.00275 }&
 {\scriptsize 0.00258 }\tabularnewline
 {\scriptsize 3 }&
 {\scriptsize 1.580 }&
 {\scriptsize 2.879 }&
 {\scriptsize 3 }&
 {\scriptsize 0.226 }&
 {\scriptsize 0.198 }&
 {\scriptsize 3 }&
 {\scriptsize 0.00167 }&
 {\scriptsize 0.00658 }\tabularnewline
 {\scriptsize 4 }&
 {\scriptsize 25.910 }&
 {\scriptsize 44.143 }&
 {\scriptsize 4 }&
 {\scriptsize $1.300\times10^{-4}$}&
 {\scriptsize $7.718\times10^{-05}$}&
 {\scriptsize 4 }&
 {\scriptsize 0.00155 }&
 {\scriptsize 0.00044 }\tabularnewline
 {\scriptsize 5 }&
 {\scriptsize 0.457 }&
 {\scriptsize 0.854 }&
 {\scriptsize 5 }&
 {\scriptsize 1.262 }&
 {\scriptsize 1.145 }&
 {\scriptsize 5 }&
 {\scriptsize 0.0366 }&
 {\scriptsize 0.01836 }\tabularnewline
 {\scriptsize 6 }&
 {\scriptsize 1.298 }&
 {\scriptsize 2.297 }&
 {\scriptsize 6 }&
 {\scriptsize 0.118 }&
 {\scriptsize 0.0927 }&
 {\scriptsize 6 }&
 {\scriptsize 0.00176 }&
 {\scriptsize 0.00819 }\tabularnewline
 {\scriptsize 7 }&
 {\scriptsize 2.516 }&
 {\scriptsize 4.342 }&
 {\scriptsize 7 }&
 {\scriptsize 0.0175 }&
 {\scriptsize 0.0113 }&
 {\scriptsize 7 }&
 {\scriptsize 0.00416 }&
 {\scriptsize 0.00443 }\tabularnewline
 {\scriptsize 8 }&
 {\scriptsize 3.294 }&
 {\scriptsize 5.933 }&
 {\scriptsize 8 }&
 {\scriptsize 0.0834 }&
 {\scriptsize 0.0691 }&
 {\scriptsize 8 }&
 {\scriptsize 0.00343 }&
 {\scriptsize 0.00332 }\tabularnewline
 {\scriptsize 9 }&
 {\scriptsize 2.371 }&
 {\scriptsize 4.592  }&
 {\scriptsize 9 }&
 {\scriptsize 0.466 }&
 {\scriptsize 0.464  }&
 {\scriptsize 9 }&
 {\scriptsize 0.00275 }&
 {\scriptsize 0.00417  }\tabularnewline
\hline
\multicolumn{3}{|c||}{{\scriptsize $h^{0}\rightarrow b+\overline{b}$}}&
\multicolumn{3}{c||}{{\scriptsize $H^{+}\rightarrow t+\overline{b}$}}&
\multicolumn{3}{c|}{}\tabularnewline
\cline{1-3} \cline{4-6} 
{\scriptsize SPS }&
 {\scriptsize $\Gamma_{I}$}&
 {\scriptsize $\Gamma_{F}$}&
{\scriptsize SPS }&
 {\scriptsize $\Gamma_{I}$}&
 {\scriptsize $\Gamma_{F}$}&
&
&
\tabularnewline
{\scriptsize 1 }&
 {\scriptsize 0.00482 }&
 {\scriptsize 0.00685 }&
{\scriptsize 1 }&
 {\scriptsize 0.835 }&
 {\scriptsize 1.463 }&
&
&
\tabularnewline
 {\scriptsize 2 }&
 {\scriptsize 0.00405 }&
 {\scriptsize 0.00575 }&
 {\scriptsize 2 }&
 {\scriptsize 3.981 }&
 {\scriptsize 8.1306 }&
&
&
\tabularnewline
 {\scriptsize 3 }&
 {\scriptsize 0.00445 }&
 {\scriptsize 0.00629 }&
 {\scriptsize 3 }&
 {\scriptsize 1.473 }&
 {\scriptsize 2.696 }&
&
&
\tabularnewline
 {\scriptsize 4 }&
 {\scriptsize 0.00413 }&
 {\scriptsize 0.00733 }&
 {\scriptsize 4 }&
 {\scriptsize 14.635 }&
 {\scriptsize 26.815 }&
&
&
\tabularnewline
 {\scriptsize 5 }&
 {\scriptsize 0.00449 }&
 {\scriptsize 0.00626 }&
 {\scriptsize 5 }&
 {\scriptsize 1.697 }&
 {\scriptsize 2.229 }&
&
&
\tabularnewline
 {\scriptsize 6 }&
 {\scriptsize 0.00462 }&
 {\scriptsize 0.00652 }&
 {\scriptsize 6 }&
 {\scriptsize 1.0817 }&
 {\scriptsize 1.930 }&
&
&
\tabularnewline
 {\scriptsize 7 }&
 {\scriptsize 0.00509 }&
 {\scriptsize 0.00691 }&
 {\scriptsize 7 }&
 {\scriptsize 1.577 }&
 {\scriptsize 2.938 }&
&
&
\tabularnewline
 {\scriptsize 8 }&
 {\scriptsize 0.00456 }&
 {\scriptsize 0.00632 }&
 {\scriptsize 8 }&
 {\scriptsize 2.531 }&
 {\scriptsize 4.898 }&
&
&
\tabularnewline
 {\scriptsize 9 }&
 {\scriptsize 0.00414 }&
 {\scriptsize 0.00586  }&
 {\scriptsize 9 }&
 {\scriptsize 2.484 }&
 {\scriptsize 4.801  }&
&
&
\tabularnewline
\hline
\end{tabular}{\scriptsize \par}
\caption{Higgs decay widths in ISAJET (I) and FDCSUSYDecay (F)\label{tblhiggsdecay}}
\end{table}

In the following we present a detailed disscussion for the second kind of 
discrepancy. There are 3 kinds of decay modes with different decay equations 
between ISAJET and FDCSUSYDecay. They are: 
\begin{enumerate}
\item A chargino decays into a left-sstrange and a c-quark ($\tilde{\chi}_{i}^{+}\rightarrow\bar{\tilde{s}}_{L}+c$); 
\item The neutral pseudoscalar Higgs decays into 2 different staus ($A^{0}\rightarrow\tilde{\tau}_{1}^{-}+\tilde{\tau}_{2}^{-}$); 
\item The neutral pseudoscalar Higgs decays into a chargino pair ($A^{0}\rightarrow\tilde{\chi}_{i}^{-}+\tilde{\chi}_{i}^{+}$,
$i=1,2$).
\end{enumerate}
The numerical results of decay widths at SPS 1 to 9 are shown in 
Tab.\ref{decaywidth2} and it is very clear that there are large discrepancies
in these decay channels.
\begin{table}
\centering
\begin{tabular}{|c|c|c||c|c|c|}
\hline 
\multicolumn{3}{|c||}{$A^{0}\rightarrow\tilde{\tau}_{1}^{-}+\tilde{\tau}_{2}^{+}$}&
\multicolumn{3}{c|}{$A^{0}\rightarrow\chi_{1}^{-}+\chi_{1}^{+}$}\tabularnewline
\hline
SPS &
 $\Gamma_{I}$&
 $\Gamma_{F}$&
SPS &
 $\Gamma_{I}$&
 $\Gamma_{F}$\tabularnewline
\hline
1&
 0.00442&
 0.00594 &
1 &
0.0519 &
 0.258 \tabularnewline
 2 &
 - &
 -&
 2 &
1.494 &
 1.642 \tabularnewline
 3&
 0.00351&
 0.00377 &
 3 &
 - &
 -\tabularnewline
 4 &
 - &
 - &
 4 &
 - &
 - \tabularnewline
 5&
 0.0254&
 0.0305 &
 5 &
0.127 &
 0.223 \tabularnewline
 6&
 0.00111&
 0.00132 &
 6 &
0.0294 &
 0.223 \tabularnewline
 7&
 $2.76\times10^{-5}$&
 $2.71\times10^{-5}$&
 7 &
 - &
 - \tabularnewline
 8&
 $3.85\times10^{-6}$&
 $3.34\times10^{-6}$&
 8 &
0.0177 &
 0.213 \tabularnewline
 9&
 0.0367&
 0.130  &
 9 &
0.0992 &
 0.116 \tabularnewline
\hline
\multicolumn{3}{|c||}{$\chi_{2}^{+}\rightarrow\overline{\tilde{s}}_{L}+c$}&
\multicolumn{3}{c|}{$A^{0}\rightarrow\chi_{2}^{-}+\chi_{2}^{+}$}\tabularnewline
\hline
SPS &
 $\Gamma_{I}$&
 $\Gamma_{F}$&
SPS &
 $\Gamma_{I}$&
 $\Gamma_{F}$\tabularnewline
\hline
5&
 $1.86\times10^{-5}$&
 $1.7\times10^{-4}$ &
2 &
 0.354 &
 0.550  \tabularnewline
\hline
\end{tabular}
\caption{Decay width from ISAJET (I) and FDCSUSYDecay (F)\label{decaywidth2}}
\end{table} 

In the following, we briefly present the definition of the particles, the 
resulting mixing matrices and the related Feynman rules. 
The mass matrix for stau in the $\tilde{\tau}_{L}-\tilde{\tau}_{R}$
basis is \begin{eqnarray*}
M_{\tilde{\tau}} & = & \left(\begin{array}{cc}
\tilde{M}_{\tilde{\tau}_{L}}^{2} & m_{\tau}(A_{\tau}-\mu\tan\beta)\\
m_{\tau}(A_{\tau}-\mu\tan\beta) & \tilde{M}_{\tilde{\tau}_{R}}^{2}\end{array}\right)\end{eqnarray*}
\begin{equation}
\tilde{M}_{\tilde{\tau}_{L}}^{2}=m_{\tau}^{2}+\tilde{m}_{\tilde{\tau}_{1}}^{2}+m_{z}^{2}\cos2\beta(-\frac{1}{2}+\sin^{2}\theta_{W}),\end{equation}
\begin{equation}
\tilde{M}_{\tilde{\tau}_{R}}^{2}=m_{\tau}^{2}+\tilde{m}_{\tilde{\tau}_{2}}^{2}-m_{z}^{2}\cos2\beta\sin^{2}\theta_{W},\end{equation}
where $\tilde{m}^2_{\tilde{\tau}_{1}}$, $\tilde{m}^2_{\tilde{\tau}_{2}}$
and $A_{\tau}$ are parameters in soft SUSY breaking term. The stau mass matrix is
diagonalized by a unitary matrix $U^{\tilde{\tau}}$
\begin{eqnarray}
(U^{\tilde{\tau}})^{-1}M_{\tilde{\tau}}U^{\tilde{\tau}}=\left(\begin{array}{cc}
m_{\tilde{\tau}_{1}}^{2} & 0\\
0 & m_{\tilde{\tau}_{2}}^{2}\\
\end{array}\right),\hspace{3em}(m_{\tilde{\tau}_{2}}>m_{\tilde{\tau}_{1}}),\end{eqnarray}
 and \begin{equation}
\left(\begin{array}{c}
\tilde{\tau_{L}}^-\\
\tilde{\tau_{R}}^-\end{array}\right)=U^{\tilde{\tau}}\left(\begin{array}{c}
\tilde{\tau_{1}}^-\\
\tilde{\tau_{2}}^-\end{array}\right),\hspace{3em}U^{\tilde{\tau}}=\left(\begin{array}{cc}
{\cos{\theta_{\tau}}} & {\sin{\theta_{\tau}}}\\
-{\sin{\theta_{\tau}}} & {\cos{\theta_{\tau}}}\end{array}\right).\end{equation}
We adopt the positively charged charginos as particles and define the physical states with mass $m_{\tilde{\chi_i}}$ as
\begin{equation}
\Psi(\tilde{\chi}_{i}^{+})=\left(\begin{array}{c}
\overline{\lambda_{iR}^{-}}\\
\lambda_{iL}^{+}\end{array}\right),\hspace{2em}\Psi(\tilde{\chi}_{i}^{-})\equiv\Psi(\tilde{\chi}_{i}^{+})^{C}=\left(\begin{array}{c}
\overline{\lambda_{iL}^{+}}\\
\lambda_{iR}^{-}\end{array}\right)\end{equation}
 where \begin{displaymath} 
\left(\begin{array}{c}
\lambda^{-}\\
\tilde{H}_{1}^{-}\end{array}\right)=U\left(\begin{array}{c}
\lambda_{1R}^{-}\\
\lambda_{2R}^{-}\end{array}\right),\hspace{2em}\left(\begin{array}{c}
\lambda^{+}\\
\tilde{H}_{2}^{+}\end{array}\right)=V\left(\begin{array}{c}
\lambda_{1L}^{+}\\
\lambda_{2L}^{+}\end{array}\right).\end{displaymath}
 The $\lambda^{\pm}$ denote winos and $\tilde{H}_{1}^{-}$ and $\tilde{H}_{2}^{+}$
denote charged higgsinos. The unitary matrices $U$ and $V$ are given as \begin{eqnarray} \label{charmatr}
U=\left(\begin{array}{cc}
\cos\theta_{R} & \sin\theta_{R}\\
-\sin\theta_{R} & \cos\theta_{R}\end{array}\right)\left(\begin{array}{cc}
\eta_{1} & 0\\
0 & \eta_{2}\end{array}\right),\hspace{2em}V=\left(\begin{array}{cc}
\cos\theta_{L} & \sin\theta_{L}\\
-\sin\theta_{L} & \cos\theta_{L}\end{array}\right),\end{eqnarray}
 The chargino mass matrix is \begin{equation}
M_{\tilde{\chi}}=\left(\begin{array}{cc}
M_{2} & \sqrt{2}m_{z}\cos\theta_{W}\sin\beta\\
\sqrt{2}m_{z}\cos\theta_{W}\cos\beta & \mu\end{array}\right).\end{equation}
 It can be diagonalized by unitary matrices in (\ref{charmatr})
 \begin{equation}
U^{-1}M_{\tilde{\chi}}V=\left(\begin{array}{cc}
\eta_{1}m_{\tilde{c}_{1}} & 0\\
0 & \eta_{2}m_{\tilde{c}_{2}}\end{array}\right),\hspace{2em}(|m_{\tilde{c}_{2}}|>|m_{\tilde{c}_{1}}|),\end{equation}
 where $\eta_{1}=$sign$(m_{\tilde{c}_{1}})$ and $\eta_{2}=$sign$(m_{\tilde{c}_{2}})$.
 The chargino mass $m_{\tilde{\chi}_{i}}$ then will be 
 \begin{equation}
m_{\tilde{\chi}_{1}}=\eta_{1}m_{\tilde{c}_{1}},\hspace{2em}m_{\tilde{\chi}_{2}}=\eta_{2}m_{\tilde{c}_{2}}.\end{equation}

\begin{figure}[H]
\centering\includegraphics[%
  bb=207bp 447bp 435bp 770bp,
  scale=0.7]{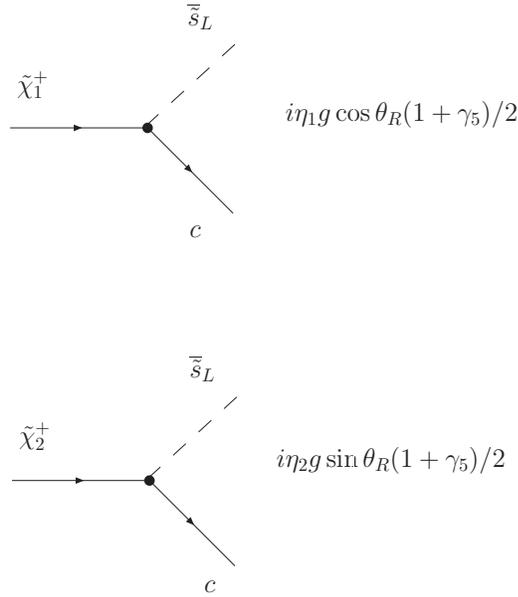}
\caption{Vertices of $\tilde{\chi}_{i}^{+}\bar{\tilde{s}}_{L}c$ in 
FDC\label{figCHI}}
\end{figure}

\begin{figure}[H]
\centering
\includegraphics[%
  bb=83bp 472bp 469bp 770bp,
  scale=0.8]{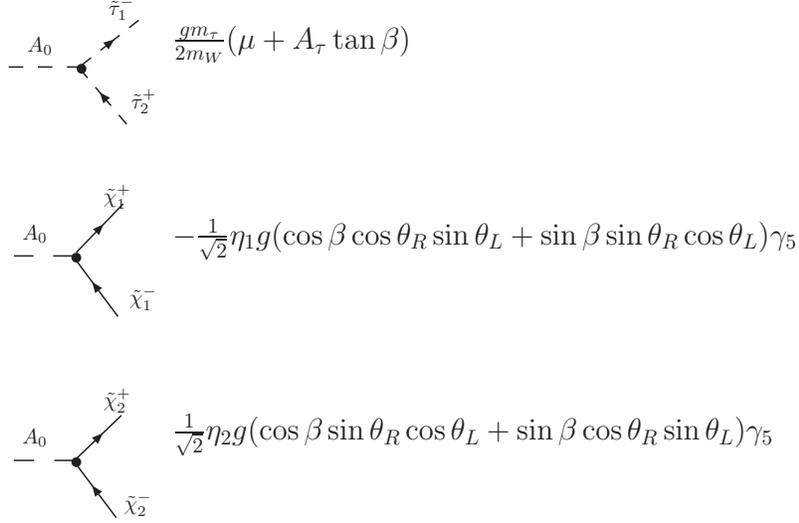}
\caption{Vertices of $A_{0}\tilde{\tau}_{1}^{-}\tilde{\tau}_{2}^{-}$
and $A_{0}\tilde{\chi}_{i}^{-}\tilde{\chi}_{i}^{+}$\label{figHA0}}
\end{figure}

The related Feynman rules used in our package are given in Fig.\ref{figCHI}  
and \ref{figHA0}, which have been carefully checked with those in Ref. 
\cite{Kuroda:1999ks}.

and decay width equations are listed as: 
\begin{eqnarray}
\Gamma(\tilde{\chi}_{1}^{+}\rightarrow\overline{\tilde{s}}_{L}+c) & = & \frac{3g^{2}|P_{cm}|}{16\pi m_{\tilde{\chi}_{1}}^{2}}(m_{c}^{2}-m_{\tilde{s}_{L}}^{2}+m_{\tilde{\chi}_{1}}^{2})\cos^2\theta_R\\
\Gamma(\tilde{\chi}_{2}^{+}\rightarrow\overline{\tilde{s}}_{L}+c) & = & \frac{3g^{2}|P_{cm}|}{16\pi m_{\tilde{\chi}_{2}}^{2}}(m_{c}^{2}-m_{\tilde{s}_{L}}^{2}+m_{\tilde{\chi}_{2}}^{2})\sin^2\theta_R\\
\Gamma(A^{0}\rightarrow\tilde{\tau}_{1}^{-}+\tilde{\tau}_{2}^{+}) & = & \frac{g^{2}m_{\tau}^{2}|P_{cm}|}{32\pi m_{A^{0}}^{2}m_{W}^{2}}(\mu+A_{\tau}\tan\beta)^{2}\\
 \Gamma(A^{0}\rightarrow\tilde{\chi}_{1}^{-}+\tilde{\chi}_{1}^{+}) & = & \frac{g^{2}|P_{cm}|}{8\pi}(\cos\beta\cos\theta_R\sin\theta_L+\sin\beta\sin\theta_R\cos\theta_L)^{2}\\
\Gamma(A^{0}\rightarrow\tilde{\chi}_{2}^{-}+\tilde{\chi}_{2}^{+}) & = & \frac{g^{2}|P_{cm}|}{8\pi}(\cos\beta\sin\theta_R\cos\theta_L+\sin\beta\cos\theta_R\sin\theta_L)^{2} 
 \end{eqnarray}
where $|P_{cm}|$ is the momentum of one of the final particle in the rest
frame of the initial particle\begin{eqnarray*}
|P_{cm}| & = & \frac{\sqrt{\left[M^{2}-(m_{1}+m_{2})^{2}\right]\left[M^{2}-(m_{1}-m_{2})^{2}\right]}}{2M}\end{eqnarray*}
\\
where $M$ is the mass of the initial particle and $m_{1},m_{2}$
is the masses of the final particles.
\section{Summary}

We have presented the FORTRAN program package FDCSUSYDecay, which
is used to calculate all the possible SUSY particle and Higgs
2-body decays in the MSSM model at tree-level. 
The package with needed 3-body channels can be generated by
FDC when the MSSM parameters are fixed.  One of the attractive
features of this package is that all the FORTRAN codes are generated
by FDC with full automation and it can avoid uncareful errors.
Although this package can
now only work at tree-level, the ability of
one-loop calculation in FDC are under developing. A new version of FDCSUSYDecay
with full one-loop correction will be generated in near future. 

\section{Test Run and Output\label{sec:Test-Run}}
\subsection{Test-run in high-energy scale mode with parameters of Snowmass benchmark point 1}
The screen output when running \emph{fsdecayhigh} is as following:
{\footnotesize
\begin{verbatim}
**************************************
* FDCSUSYDecay 1.00                  *
**************************************
Enter Snowmass point (1-9):
OR Enter 0 to specify scenario and parameters manually:
1
Enter output file name (in single quotes):
'testrun1.dat'

\end{verbatim}
}

The decay table is stored in file \emph{testrun1.dat} and reads
like (partly listed):

{\footnotesize
\begin{verbatim}
#               =====================================
#                DECAY TABLE CREATED BY FDCSUSYDECAY
#                DATE: 11/ 4/2006    16:43:46
#               =====================================
#
#       PDG         WIDTH(GeV)
DECAY   2000006     9.00247665E+00   # ~t_2 decays              
#    BR                NDA   ID1       ID2
     2.25878090E-01    2          23   1000006   # ~t_2 -> Z0  ~t_1         
     2.21288383E-01    2           5   1000037   # ~t_2 -> b  ~chi+_2       
     1.90569923E-01    2           6   1000035   # ~t_2 -> t  ~chi0_4       
     1.61424758E-01    2           5   1000024   # ~t_2 -> b  ~chi+_1       
     6.33183903E-02    2           6   1000023   # ~t_2 -> t  ~chi0_2       
     3.98983142E-02    2           6   1000025   # ~t_2 -> t  ~chi0_3       
     3.91332313E-02    2          24   1000005   # ~t_2 -> W+  ~b_1         
     3.31751880E-02    2          25   1000006   # ~t_2 -> h0  ~t_1         
     2.53137219E-02    2           6   1000022   # ~t_2 -> t  ~chi0_1       
 
 
#       PDG         WIDTH(GeV)
DECAY   1000021     6.33793298E+00   # ~g decays                
#    BR                NDA   ID1       ID2
     1.19224123E-01    2           5  -1000005   # ~g -> b  ~b_1bar         
     1.19224123E-01    2          -5   1000005   # ~g -> bar  ~b_1          
     5.10661138E-02    2           2  -2000002   # ~g -> u  ~u_Rbar         
     5.10661138E-02    2          -2   2000002   # ~g -> ubar  ~u_R         
     5.10542957E-02    2           4  -2000004   # ~g -> c  ~c_Rbar         
     5.10542957E-02    2          -4   2000004   # ~g -> cbar  ~c_R         
     5.08960759E-02    2           1  -2000001   # ~g -> d  ~d_Rbar         
     5.08960759E-02    2          -1   2000001   # ~g -> dbar  ~d_R         
     5.08958197E-02    2           3  -2000003   # ~g -> s  ~s_Rbar         
     5.08958197E-02    2          -3   2000003   # ~g -> sbar  ~s_R         
     3.90571495E-02    2           6  -1000006   # ~g -> t  ~t_1bar         
     3.90571495E-02    2          -6   1000006   # ~g -> tbar  ~t_1         
     3.89890226E-02    2           5  -2000005   # ~g -> b  ~b_2bar         
     3.89890226E-02    2          -5   2000005   # ~g -> bar  ~b_2          
     2.81103874E-02    2           2  -1000002   # ~g -> u  ~u_Lbar         
     2.81103874E-02    2          -2   1000002   # ~g -> ubar  ~u_L         
     2.80978313E-02    2           4  -1000004   # ~g -> c  ~c_Lbar         
     2.80978313E-02    2          -4   1000004   # ~g -> cbar  ~c_L         
     2.13047293E-02    2           1  -1000001   # ~g -> d  ~d_Lbar         
     2.13047293E-02    2          -1   1000001   # ~g -> dbar  ~d_L         
     2.13044513E-02    2           3  -1000003   # ~g -> s  ~s_Lbar         
     2.13044513E-02    2          -3   1000003   # ~g -> sbar  ~s_L         
 
 
#       PDG         WIDTH(GeV)
DECAY   1000003     5.00062352E+00   # ~s_L decays              
#    BR                NDA   ID1       ID2
     6.10919249E-01    2           4  -1000024   # ~s_L -> c  ~chi-_1       
     3.10762902E-01    2           3   1000023   # ~s_L -> s  ~chi0_2       
     3.89466555E-02    2           4  -1000037   # ~s_L -> c  ~chi-_2       
     2.41598639E-02    2           3   1000022   # ~s_L -> s  ~chi0_1       
     1.36997584E-02    2           3   1000035   # ~s_L -> s  ~chi0_4       
     1.51157142E-03    2           3   1000025   # ~s_L -> s  ~chi0_3       
 
 
...
\end{verbatim}
}

\subsection{Test-run in high-energy scale mode with manually inputed parameters}
The screen output when running \emph{fsdecayhigh} is as following:
{\footnotesize
\begin{verbatim}
**************************************
* FDCSUSYDecay 1.00                  *
**************************************
Enter Snowmass point (1-9):
OR Enter 0 to specify scenario and parameters manually:
0
1.mSUGRA
2.minimal Gauge-Mediated SUSY Breaking (mGMSB)
3.Anomaly-Mediated SUSY Breaking (AMSB)
Please choose scenario (1-3):
1
Enter M_0, M_(1/2), A_0, tan(BETA), sign(MU), M_t(top mass)
100.0  260.0  -100.0  12.0  1.0  175.0                                         
Enter output file name (in single quotes):
'testrun2.dat'
\end{verbatim}
}

The decay table is stored in file \emph{testrun2.dat} and reads
like (partly listed):
{\footnotesize
\begin{verbatim}
#               =====================================
#                DECAY TABLE CREATED BY FDCSUSYDECAY
#                DATE: 11/ 4/2006    16:50:38
#               =====================================
#
#       PDG         WIDTH(GeV)
DECAY   1000021     6.87489451E+00   # ~g decays                
#    BR                NDA   ID1       ID2
     1.22817522E-01    2           5  -1000005   # ~g -> b  ~b_1bar         
     1.22817522E-01    2          -5   1000005   # ~g -> bar  ~b_1          
     5.00789610E-02    2           1  -2000001   # ~g -> d  ~d_Rbar         
     5.00789610E-02    2          -1   2000001   # ~g -> dbar  ~d_R         
     5.00787337E-02    2           3  -2000003   # ~g -> s  ~s_Rbar         
     5.00787337E-02    2          -3   2000003   # ~g -> sbar  ~s_R         
     4.98299423E-02    2           2  -2000002   # ~g -> u  ~u_Rbar         
     4.98299423E-02    2          -2   2000002   # ~g -> ubar  ~u_R         
     4.98194437E-02    2           4  -2000004   # ~g -> c  ~c_Rbar         
     4.98194437E-02    2          -4   2000004   # ~g -> cbar  ~c_R         
     4.34327716E-02    2           6  -1000006   # ~g -> t  ~t_1bar         
     4.34327716E-02    2          -6   1000006   # ~g -> tbar  ~t_1         
     3.66751775E-02    2           5  -2000005   # ~g -> b  ~b_2bar         
     3.66751775E-02    2          -5   2000005   # ~g -> bar  ~b_2          
     2.74132411E-02    2           2  -1000002   # ~g -> u  ~u_Lbar         
     2.74132411E-02    2          -2   1000002   # ~g -> ubar  ~u_L         
     2.74020784E-02    2           4  -1000004   # ~g -> c  ~c_Lbar         
     2.74020784E-02    2          -4   1000004   # ~g -> cbar  ~c_L         
     2.12261878E-02    2           1  -1000001   # ~g -> d  ~d_Lbar         
     2.12261878E-02    2          -1   1000001   # ~g -> dbar  ~d_L         
     2.12259409E-02    2           3  -1000003   # ~g -> s  ~s_Lbar         
     2.12259409E-02    2          -3   1000003   # ~g -> sbar  ~s_L         
 
 
#       PDG         WIDTH(GeV)
DECAY   2000006     9.48857394E+00   # ~t_2 decays              
#    BR                NDA   ID1       ID2
     2.19460985E-01    2           5   1000037   # ~t_2 -> b  ~chi+_2       
     2.12621918E-01    2          23   1000006   # ~t_2 -> Z0  ~t_1         
     1.97473598E-01    2           6   1000035   # ~t_2 -> t  ~chi0_4       
     1.65481446E-01    2           5   1000024   # ~t_2 -> b  ~chi+_1       
     6.29074190E-02    2           6   1000023   # ~t_2 -> t  ~chi0_2       
     4.22399775E-02    2           6   1000025   # ~t_2 -> t  ~chi0_3       
     4.13479750E-02    2          24   1000005   # ~t_2 -> W+  ~b_1         
     3.30474133E-02    2          25   1000006   # ~t_2 -> h0  ~t_1         
     2.54192679E-02    2           6   1000022   # ~t_2 -> t  ~chi0_1       
 
 
#       PDG         WIDTH(GeV)
DECAY   1000003     5.17658881E+00   # ~s_L decays              
#    BR                NDA   ID1       ID2
     6.11686309E-01    2           4  -1000024   # ~s_L -> c  ~chi-_1       
     3.11939618E-01    2           3   1000023   # ~s_L -> s  ~chi0_2       
     3.86054484E-02    2           4  -1000037   # ~s_L -> c  ~chi-_2       
     2.27239525E-02    2           3   1000022   # ~s_L -> s  ~chi0_1       
     1.35532277E-02    2           3   1000035   # ~s_L -> s  ~chi0_4       
     1.49144465E-03    2           3   1000025   # ~s_L -> s  ~chi0_3       
     
...
\end{verbatim}
}
\subsection{Test-run in low-energy scale mode}
The file \emph{usrpara.f} is prepared as:
{\footnotesize
\begin{verbatim}
      SUBROUTINE FUPAR()
c-----------------------------------------------------------------
c      By this subroutine, users can specify the MSSM and SM
c      parameters at low-energy scale.
c
c      **  IMPORTANT **
c        PROGRAM DO NOT CHECK THE VALIDITY OF PARAMETERS VALUE
c        YOU INPUTED.
c
c      **REMEMBER TO RE-BUILD THE PROGRAM AFTER EVERY MODIFICATION
c        BY COMMAND:
c         make lowscale
c
c-----------------------------------------------------------------
c      The MSSM parameter variables are listed below:
c      The unit is GeV
c      
c       FSSGM1     =   U(1) gaugino mass parameter           
c       FSSGM2     =   U(2) gaugino mass parameter
c       FSSGM3     =   U(3) gaugino mass parameter    
c       FSSHMU     =   Higgsino mass parameter    
c       FSSTANB    =   ratio of Higgs vacuum expectation
c       FSSMHA     =   neutral pseudoscalar Higgs(A0) mass
c       FSSAAL     =   trilinear coupling for tau lepton
c       FSSAAT     =   trilinear coupling for t-quark
c       FSSAAB     =   trilinear coupling for b-quark
c       FSSSBEL    =   SUSY-breaking mass parameter for left s-electron
c       FSSSBER    =   SUSY-breaking mass parameter for right s-electron     
c       FSSSBMUL   =   SUSY-breaking mass parameter for left s-muon
c       FSSSBMUR   =   SUSY-breaking mass parameter for right s-muon      
c       FSSSBTAU1  =   SUSY-breaking mass parameter for left s-tau       
c       FSSSBTAU2  =   SUSY-breaking mass parameter for right s-tau       
c       FSSSBUPL   =   SUSY-breaking mass parameter for left s-up      
c       FSSSBUPR   =   SUSY-breaking mass parameter for right s-up
c       FSSSBCHL   =   SUSY-breaking mass parameter for left s-charm
c       FSSSBCHR   =   SUSY-breaking mass parameter for right s-charm
c       FSSSBDNR   =   SUSY-breaking mass parameter for right s-down
c       FSSSBSTR   =   SUSY-breaking mass parameter for right s-strange
c       FSSSBBT2   =   SUSY-breaking mass parameter for heavy s-bottom
c       FSSSBTP1   =   SUSY-breaking mass parameter for light s-top
c       FSSSBTP2   =   SUSY-breaking mass parameter for heavy s-top
c-----------------------------------------------------------------
c      The SM parameter variables are listed below:
c      The unit is GeV
c      
c       FSMME     =  electron mass
c       FSMMMU    =  muon mass
c       FSMMTAU   =  tau mass
c       FSMMUP    =  u-quark mass
c       FSMMDN    =  d-quark mass    
c       FSMMCH    =  c-quark mass    
c       FSMMST    =  s-quark mass   
c       FSMMTP    =  t-quark mass 
c       FSMMBT    =  b-quark mass    
c       FSMMZ     =  Z boson mass 
c       FSMSN2THW =  weak-mixing angle SIN^2(THETA_W)
c       FSMFNST   =  fine-structure constant ALPHA_EM
c       FSMALPS   =  strong-coupling constant ALPHA_S
c-----------------------------------------------------------------
         IMPLICIT NONE 
         
cc BLOCK FUSSPAR START         
         DOUBLE PRECISION FSSGM1, FSSGM2,FSSGM3,FSSHMU,FSSTANB,FSSMHA,
     $     FSSAAL,FSSAAT,FSSAAB,FSSSBEL,FSSSBER,FSSSBMUL,FSSSBMUR,
     $     FSSSBTAU1,FSSSBTAU2,FSSSBUPL,FSSSBUPR,FSSSBCHL,FSSSBCHR,
     $     FSSSBDNR,FSSSBSTR,FSSSBBT2,FSSSBTP1,FSSSBTP2

         COMMON /FUSSPAR/ FSSGM1, FSSGM2,FSSGM3,FSSHMU,FSSTANB,FSSMHA,
     $     FSSAAL,FSSAAT,FSSAAB,FSSSBEL,FSSSBER,FSSSBMUL,FSSSBMUR,
     $     FSSSBTAU1,FSSSBTAU2,FSSSBUPL,FSSSBUPR,FSSSBCHL,FSSSBCHR,
     $     FSSSBDNR,FSSSBSTR,FSSSBBT2,FSSSBTP1,FSSSBTP2
cc BLOCK END         

cc BLOCK FUSMPAR START         
         DOUBLE PRECISION FSMME,FSMMMU,FSMMTAU,FSMMUP,FSMMDN,FSMMCH,
     $    FSMMST,FSMMBT,FSMMTP,FSMMZ,FSMSN2THW,FSMFNST,FSMALPS

         COMMON /FUSMPAR/ FSMME,FSMMMU,FSMMTAU,FSMMUP,FSMMDN,FSMMCH,
     $    FSMMST,FSMMBT,FSMMTP,FSMMZ,FSMSN2THW,FSMFNST,FSMALPS
cc BLOCK END         

cc SET YOUR MSSM PARAMETERS HERE

         FSSGM1     = 99.6476898d0
         FSSGM2     = 192.514954d0
         FSSGM3     = 579.09314d0
         FSSHMU     = 354.201385d0
         FSSTANB    = 10.0D0
         FSSMHA     = 394.874817D0
         FSSAAL     = -253.15593d0
         FSSAAB     = -766.899109d0
         FSSAAT     =  -495.786896d0
         FSSSBEL    = 38581.48830D0
         FSSSBER    = 18547.65040D0
         FSSSBMUL   = 38581.48830D0
         FSSSBMUR   = 18547.65040D0
         FSSSBTAU1  = 38243.67190D0
         FSSSBTAU2  = 17847.84380D0
         FSSSBUPL   = 289769.750D0
         FSSSBUPR   = 271373.5620D0
         FSSSBCHL   = 289769.750D0
         FSSSBCHR   = 271373.5620D0
         FSSSBDNR   = 269598.8440D0
         FSSSBSTR   = 269598.8440D0
         FSSSBBT2   = 266551.1560D0
         FSSSBTP1   = 243069.0310D0
         FSSSBTP2   = 178026.6410D0

cc SET YOUR SM PARAMETERS HERE

         FSMME     = 5.10999991D-4
         FSMMMU    = 1.04999997D-1
         FSMMTAU   = 1.77699995
         FSMMUP    = 5.5999998D-3
         FSMMDN    = 9.89999995D-3
         FSMMCH    = 1.35000002D0
         FSMMST    = 1.99000001D-1
         FSMMTP    = 175D0
         FSMMBT    = 5D0
         FSMMZ     = 91.1699982D0
         FSMSN2THW = 0.2311678d0
         FSMFNST   = 1D0/127.970535
         FSMALPS   = 0.118170463d0
      END
\end{verbatim}
}

The screen output when running \emph{fsdecaylow} is as following:
{\footnotesize
\begin{verbatim}
**************************************
* FDCSUSYDecay 1.00                  *
**************************************
ENTER output file name (in single quotes):
'testrun3.dat'
\end{verbatim}
}

The decay table is stored in file \emph{testrun3.dat} and reads
like (partly listed):
{\footnotesize
\begin{verbatim}
#               ======================================
#                DECAY TABLE CREATED BY FDCSUSYDECAY
#                DATE: 11/ 4/2006    16:58:46
#               ======================================
###
#       PDG         WIDTH(GeV)
DECAY   2000006     9.00247595E+00   # ~t_2 decays              
#    BR                NDA   ID1       ID2
     2.25878090E-01    2          23   1000006   # ~t_2 -> Z0  ~t_1         
     2.21288384E-01    2           5   1000037   # ~t_2 -> b  ~chi+_2       
     1.90569924E-01    2           6   1000035   # ~t_2 -> t  ~chi0_4       
     1.61424752E-01    2           5   1000024   # ~t_2 -> b  ~chi+_1       
     6.33183886E-02    2           6   1000023   # ~t_2 -> t  ~chi0_2       
     3.98983144E-02    2           6   1000025   # ~t_2 -> t  ~chi0_3       
     3.91332354E-02    2          24   1000005   # ~t_2 -> W+  ~b_1         
     3.31751871E-02    2          25   1000006   # ~t_2 -> h0  ~t_1         
     2.53137244E-02    2           6   1000022   # ~t_2 -> t  ~chi0_1       
 
 
#       PDG         WIDTH(GeV)
DECAY   1000021     6.33793301E+00   # ~g decays                
#    BR                NDA   ID1       ID2
     1.19224123E-01    2           5  -1000005   # ~g -> b  ~b_1bar         
     1.19224123E-01    2          -5   1000005   # ~g -> bar  ~b_1          
     5.10661146E-02    2           2  -2000002   # ~g -> u  ~u_Rbar         
     5.10661146E-02    2          -2   2000002   # ~g -> ubar  ~u_R         
     5.10542966E-02    2           4  -2000004   # ~g -> c  ~c_Rbar         
     5.10542966E-02    2          -4   2000004   # ~g -> cbar  ~c_R         
     5.08960753E-02    2           1  -2000001   # ~g -> d  ~d_Rbar         
     5.08960753E-02    2          -1   2000001   # ~g -> dbar  ~d_R         
     5.08958191E-02    2           3  -2000003   # ~g -> s  ~s_Rbar         
     5.08958191E-02    2          -3   2000003   # ~g -> sbar  ~s_R         
     3.90571491E-02    2           6  -1000006   # ~g -> t  ~t_1bar         
     3.90571491E-02    2          -6   1000006   # ~g -> tbar  ~t_1         
     3.89890229E-02    2           5  -2000005   # ~g -> b  ~b_2bar         
     3.89890229E-02    2          -5   2000005   # ~g -> bar  ~b_2          
     2.81103873E-02    2           2  -1000002   # ~g -> u  ~u_Lbar         
     2.81103873E-02    2          -2   1000002   # ~g -> ubar  ~u_L         
     2.80978312E-02    2           4  -1000004   # ~g -> c  ~c_Lbar         
     2.80978312E-02    2          -4   1000004   # ~g -> cbar  ~c_L         
     2.13047294E-02    2           1  -1000001   # ~g -> d  ~d_Lbar         
     2.13047294E-02    2          -1   1000001   # ~g -> dbar  ~d_L         
     2.13044514E-02    2           3  -1000003   # ~g -> s  ~s_Lbar         
     2.13044514E-02    2          -3   1000003   # ~g -> sbar  ~s_L         
 
 
#       PDG         WIDTH(GeV)
DECAY   1000003     5.00062300E+00   # ~s_L decays              
#    BR                NDA   ID1       ID2
     6.10919249E-01    2           4  -1000024   # ~s_L -> c  ~chi-_1       
     3.10762900E-01    2           3   1000023   # ~s_L -> s  ~chi0_2       
     3.89466552E-02    2           4  -1000037   # ~s_L -> c  ~chi-_2       
     2.41598663E-02    2           3   1000022   # ~s_L -> s  ~chi0_1       
     1.36997584E-02    2           3   1000035   # ~s_L -> s  ~chi0_4       
     1.51157143E-03    2           3   1000025   # ~s_L -> s  ~chi0_3       
...
\end{verbatim}
}

\end{document}